\documentclass{aa}
\usepackage{graphicx}
\usepackage{longtable}
\usepackage{latexsym}
\usepackage{amssymb}
\usepackage{lscape}
\usepackage{psfig}
\begin{document}
\title{Introducing GOLDMine: A new Galaxy Database on the WEB}

\author{Giuseppe Gavazzi \inst{1}
\and Alessandro Boselli \inst{2}
\and Alessandro Donati \inst{1}
\and Paolo Franzetti \inst{3,1}
\and Marco Scodeggio \inst{3}
}

   
\institute{
Universit\`a degli Studi di Milano-Bicocca, Piazza delle scienze 3, 20126, Milano, Italy\\
\email {giuseppe.gavazzi@mib.infn.it alessandro.donati@mib.infn.it}
\and
Laboratoire d'Astrophysique de Marseille, Traverse du Siphon, F-13376, Marseille
Cedex 12, France\\
\email {alessandro.boselli@oamp.fr}
\and
Istituto di Astrofisica Spaziale e Fisica Cosmica - CNR - via Bassini 15, 20133, Milano, Italy
\email {paolo@mi.iasf.cnr.it marcos@mi.iasf.cnr.it}
}
\date{}

\abstract{The new World Wide Web site "GOLDMine" (Galaxy On Line Database Milano Network) (http://goldmine.mib.infn.it)
contains a multiwavelength data-base of an optically selected sample of 3267 galaxies in the  Virgo cluster and in the  Coma
Supercluster. It is designed for professional astronomers who wish to find
 data and images for these galaxies.
 Data, gathered in 15 years of observational campaigns by the authors or taken from the literature
include  general parameters (catalogue names, celestial coordinates, morphological type, 
recessional velocity etc.); multiwavelength  continuum photometry (total UV, U, B, V, J, H, K, FIR and radio magnitudes/flux densities);
line photometry (HI, H$_2$, H$\alpha$);  dynamical parameters (rotational velocity from the HI and H$\alpha$
lines, velocity dispersion) and structural parameters (light concentration index,
effective radius and brightness, asymptotic magnitude) in the optical (B and V) and Near Infrared (H or K)
bands.\\
Images include finding charts, optical (B and V), H$\alpha$, Near Infrared (H and/or K) and true color RGB frames (when available).
Radial light profiles obtained from the B, V, H or K band images are also available.
Integrated optical spectra along with broad Spectral Energy Distributions (SED)
from the UV to the radio domain are given.
All images can be obtained in JPG format, but the original (reduced) FITS images can be downloaded as well. 
The database will be updated regularly and will be extended to other local clusters and superclusters.
Astronomers who wish to have their images included in GOLDMine are strongly encouraged to send us their material.   
}

\maketitle

%

\section{Introduction}

Galaxies come in many different forms
and sizes, but they can be broadly divided into two main species: {\it
Spirals}, with a flattened, disk-like shape, blue colors, much gas and
dust, and a widespread star formation activity that results in the
presence within them of many young stars and {\it Elliptical}, with a
spheroidal shape, red colors, little or no gas and dust, and no star
formation activity, thus containing exclusively old stars. 
While these differences are well documented, and have been
extensively studied over the last 70 years, astronomers are not yet able
to convincingly explain the origin of such a diversity. In fact one of
today's major open issues for astronomy is to obtain a plausible
reconstruction of the processes of galaxy formation and evolution. 
The task is of course enormously complicated by the sheer disparity
between cosmological and human time-scales, that makes it impossible 
for astronomers to witness galaxy evolution directly. 
One possible way to overcome this problem is to take
advantage of the ''time machine'' effect provided by the finite speed
of light. Observing today galaxies at different
distances means observing them at different epochs in the history of the
Universe, and thus with different ages.  Unfortunately
distances must be very large before age differences become significant,
and therefore unveiling the evolution of galaxies with direct
observations of very distant objects is becoming a reality only today,
with the advent of the 10m class telescopes.  Galaxies at progressively
larger look-back time (redshift), thus in a younger evolutionary stage,
will be observed in the near future, back to a small fraction of the
present age of the Universe. This will provide us with a sequence of
"fossil" galaxies, eventually disclosing the secret of their evolution,
much as fossil organisms guided entomologists tracing back the evolution
of species.  However, 15 years ago, when we initiated this project, a
direct observational approach to the problem was unconceivable, because
of the lack of proper telescopes and instrumentation. The only available
option was to attack the evolutionary enigma starting from the study of
evolved systems, i.e. the nearby galaxies. This approach is based on the
conviction that "adult" galaxies still preserve some memory of their
past, and that adequate observations would eventually help to disclose
it.\\
With this purpouse we started in 1985 an observational campaign aimed at providing the phenomenology
of local galaxies in the widest possible frequency range. We took observations
and collected data from the literature from the 2000 \AA (UV) range to the centimetric radio domain, 
spending a large effort in making the literature data as homogeneous
as possible with our own data.
GOLDMine is designed to provide access to this massive data-set
on local galaxies trough the World Wide Web. Numeric parameters as well as
the (reduced) scientific FITS images can be downloaded. 

\section{The sample selection}

\begin{figure*}[!t]
\psfig{figure=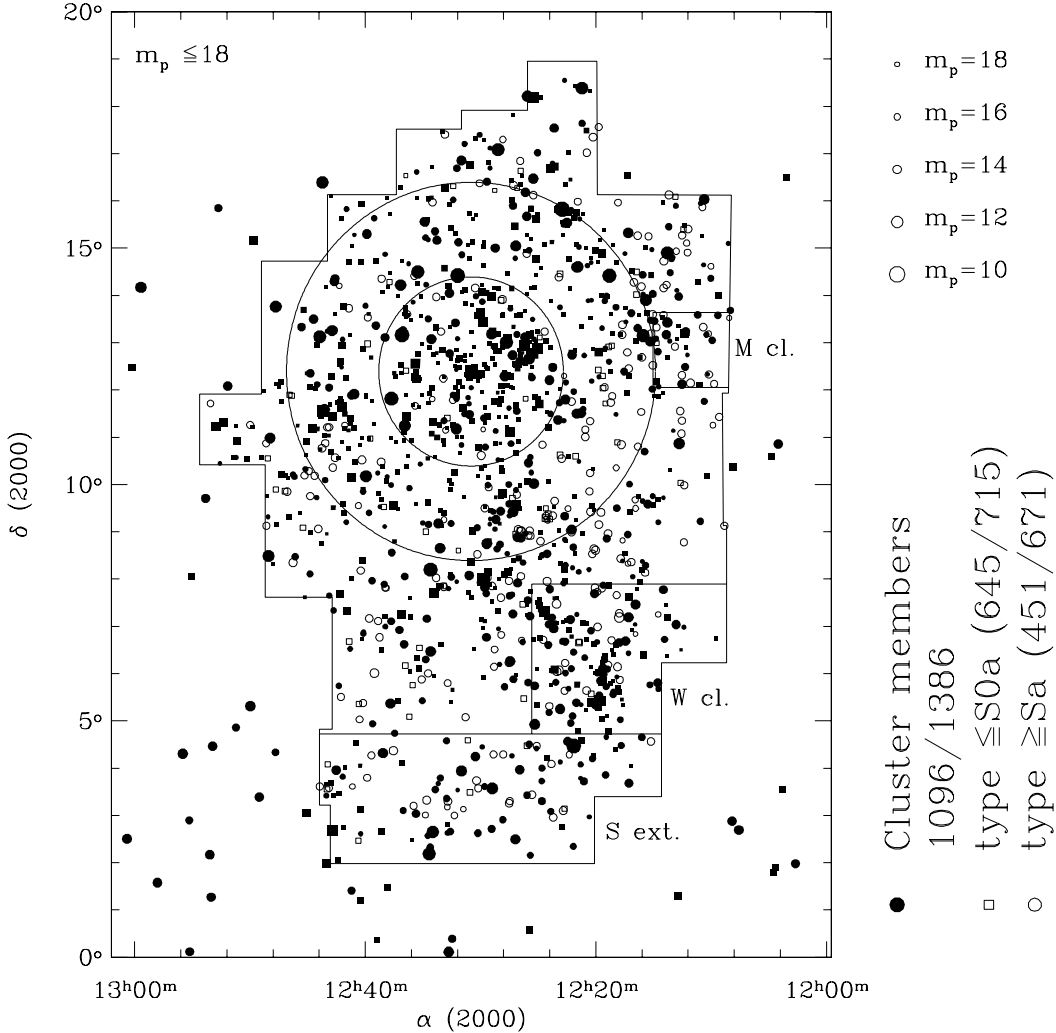,width=15cm,height=15cm}
\caption{Sky distribution of the 1386 galaxies brighter than 
$m_p \leq 18.0$ in the Virgo cluster. Possible
and spectroscopically confirmed cluster members (filled symbols) are 1096.}\label{vcccelestial}
\end{figure*}

\begin{figure*}[t!]
\psfig{figure=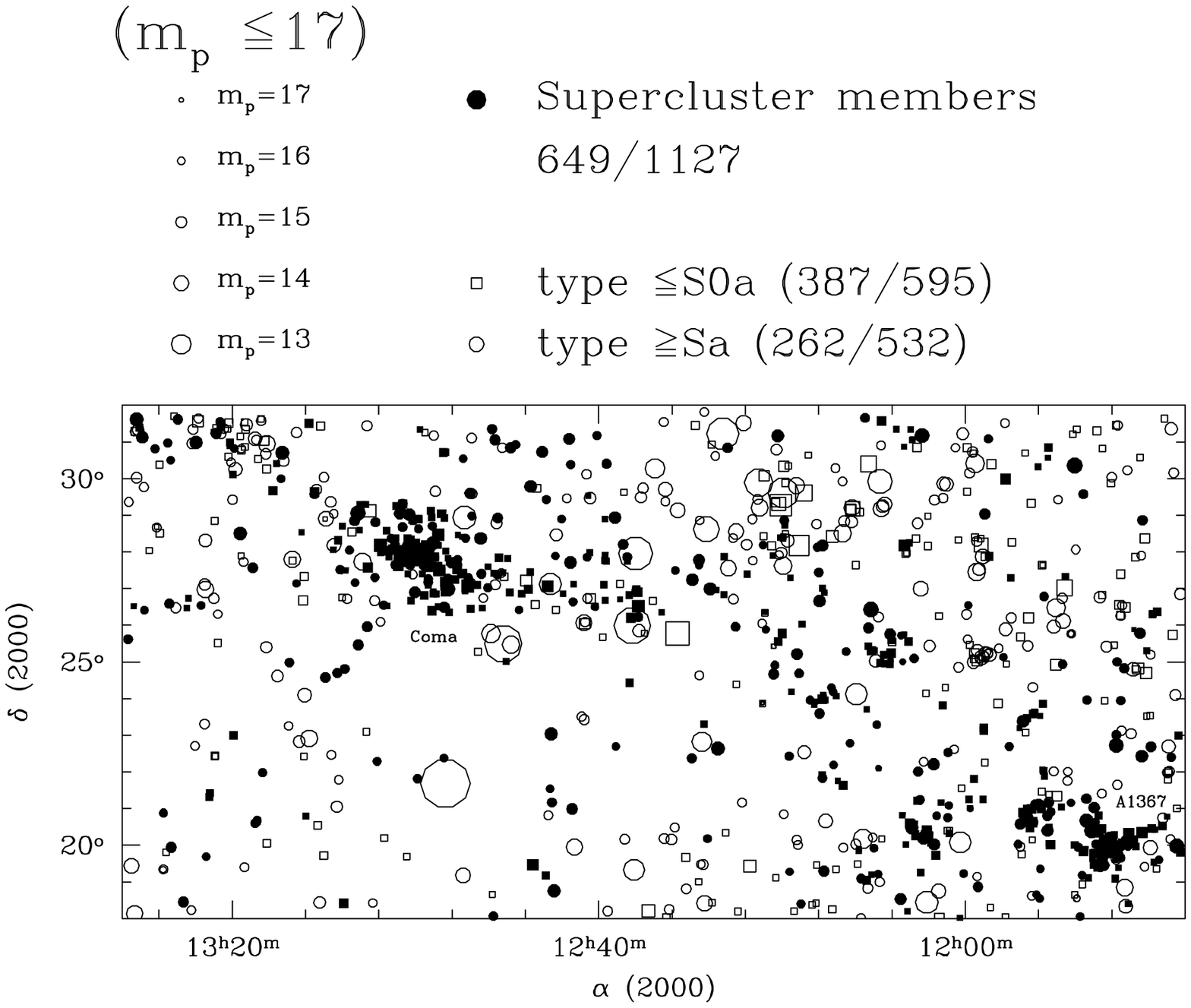,width=15cm,height=15cm}
\caption{Sky distribution of the 1127 CGCG galaxies 
in the Coma supercluster region.
Spectroscopically confirmed supercluster members (filled symbols) are 649.}\label{comascelestial}
\end{figure*}

GOLDMine is focused on two regions of the sky: the Virgo cluster and the Coma 
Supercluster\footnote{Additional 382 galaxies (with $m_p \leq 15.7$) in
the A262 (Perseus-Pisces) and Cancer clusters and in the Hercules supercluster 
(A2147, A2151, A2197, A2199) will be included shortly in GOLDMine.}. 
Among rich-clusters of galaxies, the Virgo cluster is the nearest 
to us (17 Mpc, Gavazzi et al. 1999b) (see Fig.\ref{vcccelestial}). The Coma supercluster 
contains two rich clusters (Coma and Abell 1367) corresponding to the density 
enhancements in Fig.\ref{comascelestial}. In addition, this 
region contains a filament of nearly isolated galaxies, the so called 
"Great Wall" at a distance of approximately 90 Mpc (7000 $\rm km~s^{-1}$, Gavazzi et al. 1999a) 
from us (see Fig.\ref{comascelestial}).
The studied regions contain 3 rich clusters of galaxies as well as 
isolated objects, thus providing the ideal laboratory for a comparative 
analysis of galaxies in different environments.\\
As opposed to other more extensive galaxy data-bases (eg. NED, LEDA), which include
all objects with published parameters, galaxies in GOLDMine were
selected with the criterion of optical completeness. All galaxies brighter than a threshold magnitude 
were selected in both areas. In the Coma supercluster all
galaxies brighter than $m_p=15.7$ were selected from the 
Catalogue of Galaxies and of Clusters of Galaxies (CGCG)
by Zwicky et al. (1961-1968).\footnote{The number of galaxies meeting the $m_p \leq 15.7$
criterion is 1082. However 45 galaxies listed as "pairs"
in the CGCG were split into fainter ($m_p \leq 17.0$) components. The resulting 
Coma galaxies in GOLDMine are 1127.}
The Virgo region contains all (2096) galaxies brighter than $m_p=20.0$ from the Virgo Cluster 
Catalogue (VCC) by Binggeli et al. (1985)\footnote{44 additional Virgo galaxies
with $m_p \leq 14.5$, $12^h<R.A.<13^h$, but outside the area covered by the VCC are included.}.
Obviously, due to the factor of 5 difference in distance between the 
Virgo and the Coma clusters, this selection limit results in 
dwarf galaxies being included in our database 
only for the Virgo cluster. However globally GOLDMine covers the whole range 
(4 orders of magnitude) of luminosities spanned by real galaxies.\\
Altogether GOLDMine contains 3267 galaxies, approximately 60 \% of which  
of early-type (elliptical and S0) and the remaining 40 \% 
of late-type (spirals and irregulars).\\
Extensive campaigns were carried out to observe as many as 
possible of the 3267 target galaxies through all possible observational windows, 
a task that we did not complete yet. Moreover we collected data from the
literature and we made all possible effords in trying to homogenize them 
with the ones obtained by us.


\section{Database fields}

The parameters listed in the GOLDMine database are divided into 5 categories: General, Continuum and 
Line photometry, Dynamical and Structural. All parameters are extensively described and referenced in the
section "GOLDMine glossary". See also the section "GOLDMine tutorial" to learn about using the database.\\ 
The parameters can be obtained from GOLDMine in two ways:\\
1) by querying the database for an individual galaxy name, or\\
2) by a multiple query "by parameters". In this case all galaxies in a given range of photographic magnitude,
and morphological type can be selected. The query criteria include also the availability of continuum, 
line photometry, dynamical and structural parameters.\\
General parameters include: CGCG (Zwicky et al. 1961-1968), NGC (Dreyer 1888), IC (Dreyer 1895, 1908),
VCC (Binggeli et al. 1985), UGC (Nilson 1973) designations; Celestial coordinates precessed to the year 2000; 
isophotal diameters; photographic magnitude; heliocentric recessional velocity; distance and morphological type.\\
Continuum photometry parameters consist of total magnitudes from aperture
or CCD photometry, uncorrected for extinction either internal or due to our Galaxy.
They include: UV (2000~\AA); optical (U,B,V); Near-IR (J,H,K) magnitudes; Far-IR (IRAS 60 and 100 $\mu m$);
radio continuum (1420 and 610 MHz) fluxes. \\  
Line photometry parameters include: the atomic (HI) and molecular (H$_2$) hydrogen mass (the corresponding
fluxes can be derived using the source distance); the H$\alpha$+[NII] line equivalent width
and flux.\\
Dynamical parameters include: the  width of the HI line, with a quality flag; the  width of the H$\alpha$ line and
the central velocity dispersion.\\
Structural parameters include the light concentration index (C31); the effective radius $R_e$;
the effective surface brightness $\mu_e$; the total asymptotic magnitude. These quantities are given
separately for the H, V and B bands.\\

\section{Imaging}

The novelty of GOLDMine consists of its image section, where images can be downloaded in JPG and FITS 
format. They can be accessed via the the queries "by name" or "by parameters".
Images include:\\ 
\begin{itemize}
\item{{\bf Finding Charts} from the Digitized Palomar Sky Survey. Each selected galaxy
appears at the center of the 10x10 $\rm arcmin^2$ finding chart. When multiple catalogued galaxies appear in the same 
finding chart, by clicking over their names each of them can be selected. In this case the database and image queries
are re-directed to the selected galaxy.}
\item{{\bf Broad band} images obtained in the B, V, H and K bands.}
\item{{\bf Narrow band} images in the light of H$\alpha$. Each galaxy has two H$\alpha$ images: 
the "OFF-band" one
which gives the underlying stellar continuum near H$\alpha$ and the "NET" one,
obtained by subtracting the OFF-band image from the ON-band one.}
\item{{\bf RGB} images. For few miscellaneous galaxies we combined several images to obtain "true" color
pictures (only available in JPG format). Some are combinations of H,V and B frames, other of H$\alpha$ OFF, ON and NET frames.}
\item{{\bf Light profiles}. Radial profiles of the light distribution as obtained on the available (B, V, H) images.
See Gavazzi et al. (2000).
When at least two radial profiles are available the color radial profile is also shown (only available in JPG format).}
\item{{\bf Spectroscopy}. These are optical spectra integrated over the whole surface of the
galaxy (obtained in drift-scan mode, i.e. by drifting the spectrograph slit over the galaxy extension), 
see Gavazzi et al. (2002).
Spectra were obtained in several observational campaigns carried out 
with optical telescopes (OHP 1.9m, ESO 3.6m) by Gavazzi et al. (2002). 
Spectra exist for 141 galaxies, but 80 additional spectra  will be included 
in GOLDMine shortly.}    
\item{{\bf SEDs}. These are Spectral Energy Distributions from the UV to the centimetric radio
continuum obtained from broad-band photometry (only available in JPG format). The plotted data are total fluxes (extrapolated to the
optical radii), unlike the individual aperture data given by NED. However they are given as observed, i.e.
uncorrected for extinction from our Galaxy and for internal extinction\footnote{some SEDs contain more data points
than listed in the GOLDMine database. For example the IRAS 25$ \mu m$ flux and the fluxes obtained 
by ISO (CAM and PHOT) are plotted without being listed in GOLDMine.}.}
\end{itemize}

It is our goal to provide a homogeneous set of keywords in all FITS header to characterize the data.
These will include:
\begin{itemize}
\item{Galaxy name.}
\item{Effective integration time.}
\item{Filter.}
\item{Telescope.}
\item{WCS parameters.}
\item{Photometric effective zero point (when available) 
in mag/s for broad band images or 
in $\rm erg~cm^{-2}~s^{-1}$ for H$\alpha$ images, such that $log~Flux=Zp+log~Cnts-logT_{exp}$. 
The effective zero point includes the atmospheric extinction term.}
\end{itemize}
This homogeneization is not yet complete. Its progress will be detailed in the online "GOLDMine news" section.


\section{Summary}

We undertook a 15 years research project focused on the properties of
galaxies in the local universe which brought some new insight on their
phenomenology and possibly will help disclosing their origin and evolution
(see Gavazzi 1993; Gavazzi \& Scodeggio 1996; Gavazzi et al. 1996a, 2002; Boselli et al. 2001). 
\begin{itemize}
\item{A complete, optically selected sample of 3267 galaxies representative of all morphological types
and luminosities was extensively surveyed through various observational windows from the
UV to the centimetric radio.}
\item{The surveyed regions include the Virgo cluster and the Coma supercluster.}
\item{Numeric parameters and images are made available to the international community 
through the World Wide Web site http://goldmine.mib.infn.it.}
\end{itemize} 
If your research benefits from the use of GOLDMine, 
we would appreciate the following acknowledgement in your paper: 
"This research has made use of the GOLDMine Database, operated by the Universita' degli Studi di Milano- Bicocca".
Please also cite the present paper. 

\acknowledgements 
We wish to thank all people who contributed with data to GOLDMine: Gery Bernstein, 
Christian Bonfanti, Barbara Catinella, Luis Carrasco, Fabienne Casoli, 
Alessandra Contursi,
Luca Cortese, John Dickey, Anna Gallazzi, Bianca Garilli, Jorge Iglesias-Paramo, 
Walter Jaffe, Robert Kennicutt, 
James Lequeux, Silvia Martocchi, Paola Pedotti, Daniele Pierini, Isabella Randone,
Gerry Sanvito, Ginevra Trinchieri, Richard Tuffs, Jose M. Vilchez, Stefano Zibetti. 
In particular the FOCA team (J. Donas, B. Milliard, M. Laget, 
M. Viton) is acknowledged for UV fluxes made available prior to publication.
We also wish to thank the Time Allocation Committee of many observatories where the data
here collected were taken.
GOLDMine is supported by the Physics Department of Universit\`a degli Studi di Milano-Bicocca.
A.B. acknowledges financial support from the French Programme National Galaxies (PNG). 
P.F. acknowledges support from CNR/ASI grant I/R/27/00.\\
The GOLDMine web site has been created using the Python programming language, the Apache
web server software and the MySQL database. We wish to thank the developers of these
packages for freely distributing these resources that have made possible the construction of this site.

\end{document}